\pdfoutput=1
\RequirePackage{ifpdf}
\ifpdf % We~are running pdfTeX in pdf mode
\documentclass[pdftex]{sigma}
\else
\documentclass{sigma}
\fi

\numberwithin{equation}{section}

\newtheorem{Theorem}{Theorem}[section]
\newtheorem*{Theorem*}{Theorem}

 { \theoremstyle{definition}

\newtheorem{Remark}[Theorem]{Remark} }

\DeclareMathOperator{\chord}{chord}
\DeclareMathOperator{\sign}{sign}
\usepackage{bm}

\usepackage{tikz}
\usetikzlibrary{arrows,patterns,decorations.pathmorphing,backgrounds,positioning,fit,petri,calligraphy}
\usetikzlibrary {arrows.meta}

\begin{document}
\allowdisplaybreaks

\newcommand{\arXivNumber}{2207.09606}

\renewcommand{\PaperNumber}{001}

\FirstPageHeading

\ShortArticleName{A Novel Potential Featuring Off-Center Circular Orbits}

\ArticleName{A Novel Potential Featuring Off-Center Circular\\ Orbits}

\Author{Maxim OLSHANII}

\AuthorNameForHeading{M.~Olshanii}

\Address{Department of Physics, University of Massachusetts Boston, Boston Massachusetts 02125, USA}
\Email{\href{mailto:maxim.olchanyi@umb.edu}{maxim.olchanyi@umb.edu}}
\URLaddress{\url{https://sites.google.com/view/integrability/home}}

\ArticleDates{Received October 04, 2022, in final form January 03, 2023; Published online January 07, 2023}

\Abstract{In Book 1, Proposition 7, Problem 2 of his 1687 \emph{{Philosophi\ae} Naturalis Principia Mathematica}, Isaac Newton poses and answers the following question: \emph{Let the orbit of a~particle moving in a central force field be an \emph{off-center} circle. How does the magnitude of the force depend on the position of the particle on
that circle?} In this article, we identify a~potential that can produce such a force, only at zero energy. We further map the zero-energy orbits in this potential to finite-energy free motion orbits on a sphere; such a~duality is a~particular instance of a general result by Goursat, from 1887. The map itself is an inverse stereographic projection, and this fact explains the circularity of the zero-energy orbits in the system of interest. Finally, we identify an additional integral of motion---an analogue of the Runge--Lenz vector in the Coulomb problem---that is responsible for the closeness of the zero-energy orbits in our problem.}


\Classification{37J35; 68-03}

\section{Introduction}

\begin{figure}[t]\centering
 \begin{tikzpicture}[scale=1,smooth]
 %\node at (0,0) {\includegraphics[scale=.35]{main.pdf}};
 %\draw[fill,blue] (0,0) circle [radius=.1];
 %\draw[very thin,color=gray,step=.5cm] (-7.0,-5.0) grid (7.0,5.0);
 \draw[-Stealth,thick] (-5,-0.25) -- (5,-0.25);
 \node at (5,-0.5) {$x$};
 \draw[-Stealth,thick] (-2,-4.5) -- (-2,4.5);
 \node at (-2.25,4.4) {$y$};
 \draw[thick] (0.25,-0.25) circle [radius=3.75];
 \draw[thick] (3.05,2.21) -- (-3.42,-0.93);
 \node at (-0.05,0.45) {$r$};
 \node at (-2.5,-0.7) {$r'$};
 \draw[thick] (3.05,2.21) -- (0.25,-0.25);
 \node at (1.8,0.8) {$R$};
 \draw[fill] (3.05,2.21) circle [radius=.12];
 \draw[dashed,-{Triangle[length=7pt,width=7pt]},line width=0.7mm] (3.05,2.18) -- (0.8,1.08);
 \node at (3,2.7) {$\vec{v}$};
 \draw[-{Triangle[length=7pt,width=7pt]},line width=0.7mm] (3.1,2.2) -- (2.4,2.85);
 \draw[thick] (0.75,-0.25) arc [radius=0.5, start angle=0, end angle=42];
 \node at (0.5,-0.5) {$\theta$};
 \draw[thick] (-1.5,-0.25) arc [radius=0.5, start angle=0, end angle=27];
 \node at (-1.75,-0.5) {$\phi$};
 \draw [decorate,decoration = {calligraphic brace,amplitude=12pt},ultra thick] (-3.42,-0.93) -- (3.05,2.21);
 \draw [decorate,decoration = {calligraphic brace,amplitude=9pt},ultra thick] (0.20,-1) -- (-1.95,-1);
 \node at (-0.9,-1.6) {$l$};
 \node[rotate=30] at (-0.5,1.22) {chord};
 \draw[line width=1mm] (0.25,-0.05) -- (0.25,-0.45);
 \draw[line width=1mm] (0.05,-0.25) -- (0.45,-0.25);
 \draw[line width=0.8mm] (-2,-0.25) circle [radius=0.13];
 \node at (-2.3,1) {$\mathcal{R}$};
 \node at (-2.3,-1.5) {$\mathcal{R}$};
 \draw[white,fill=white] (2.15,1.7) rectangle (2.5,1.3);
 \node at (2.3,1.5) {$\vec{F}$};
 \end{tikzpicture}

\caption{The geometry of the problem. An unknown central force $\bm{F}$, acting on a point particle of mass $m$,
generates a circular orbit with the center at $(+l,0)$ and a radius $R$.
In Book 1, Proposition 7, Problem 2 of his 1687 \emph{{Philosophi\ae} Naturalis Principia Mathematica} \cite{newton_principia_1846}
Newton shows that \emph{on the orbit}, the force must have the form $F(\theta) = -\frac{4\alpha}{r^{2}(\theta) \chord^{3}(\theta)}$. Here,
$(r, \phi)$ are the polar coordinates of the particle, and the function $\chord(\theta)$
is the length of the chord that has the particle as one of its ends and that crosses the force center. The distance $r'$ is
the complement of the distance $r$ to the full chord length. $2\mathcal{R}$ is the length of the shortest chord that
passes through the origin. The angle $\theta$ provides a convenient parametrization
for the points on the orbit.}\label{f:main}
 \end{figure}
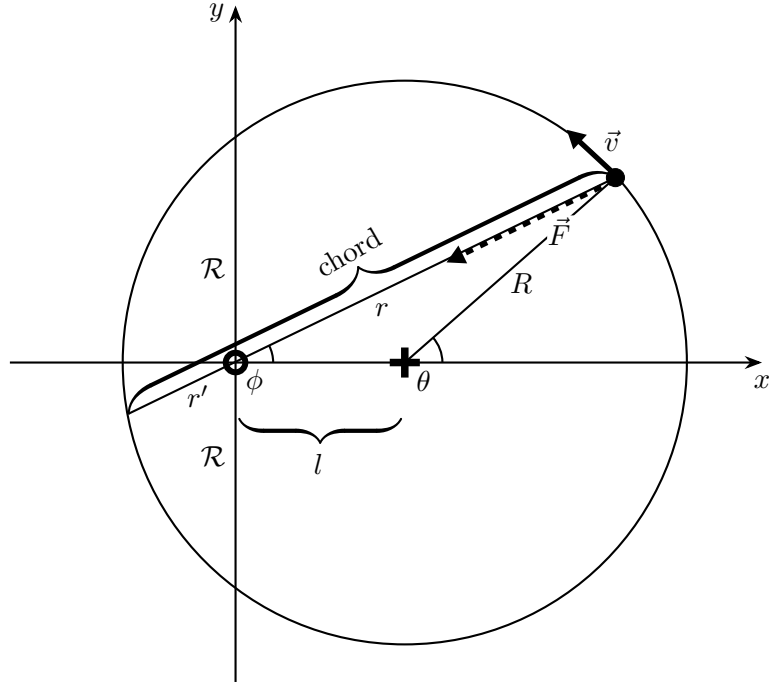

In Book 1, Proposition 7, Problem 2 of his 1687 \emph{{Philosophi\ae} Naturalis Principia Mathematica}~\cite{newton_principia_1846}, Isaac Newton poses the following question:
\begin{itemize}\itemsep=0pt
\item[] Let the orbit of a particle moving in a central force field be an \emph{off-center} circle. We assume that the orbit \emph{encloses} the potential center. How does the magnitude of the force depend on the position on
the circle?
\end{itemize}
He produces the answer that the force magnitude must depend on the position on the circle as follows:
\begin{align}
F(\theta) = -\frac{4\alpha}{r^{2}(\theta) \chord^{3}(\theta)},\label{force_720}
\end{align}
where
\begin{align}
r(\theta) = \sqrt{R^2 + l^2 + 2 Rl \cos\theta}\label{r_through_theta}
\end{align}
is the distance between the particle and the force center, and
\begin{align*}
\chord(\theta) = 2 R (R +l \cos \theta)/r(\theta)
\end{align*}
is the length of the chord of the orbital circle with the particle as one of its ends and the force center as one of its points;
both are given as functions of the particle position on the orbit, angle~$\theta$ (see Figure~\ref{f:main}). Here and below, $R$ is the radius of
the orbit, $l$ is the distance between the force center and the center of the orbit, and $\alpha$ is an arbitrary constant.

Further, in Corollary~I to \cite[Book~1, Proposition~7, Problem~2]{newton_principia_1846}, Newton looks at the particular case of orbit passing through the force center (i.e., the case of $l=R$).
In this case, ${\chord(\theta) = r(\theta)}$.
It leads to the force
\begin{align}
F_{0}(\theta) = -\frac{4\alpha}{r(\theta)^5}.\label{force_721}
\end{align}
One may further pose the question of which central potential field can produce the force \eqref{force_721} \emph{on the
orbit}. One obvious surmise is
\begin{align}
V_{0}(r) = -\frac{\alpha}{r^4}\label{potential_721}
\end{align}
everywhere.

However, to our knowledge, the problem of finding a potential that induces the force \eqref{force_720} has never been posed. In this article,
we find such a potential.

Likewise, the mathematical reason for the appearance of off-center circular orbits has been identified so far only in the limiting case
\eqref{force_721}--\eqref{potential_721}. It turns out that the potential \eqref{potential_721}, at a~given energy, is self-dual under the so-called Bohlin--Arnold--Vassiliev (BAV) duality \cite{arnold_HBNH_1990,arnold1990_1148,bohlin1911_113,horvathy2014_14042265,maclaurin_Fluxons1742}, with the dual being at a different energy and featuring a different coupling constant. In the particular case \eqref{potential_721}, the map between the two is a circular inversion. At \emph{zero energy}, the potential \eqref{potential_721} maps to a free problem, at a particular non-zero energy. Circle inversion is known to map circles and straight lines to circles or straight lines; in the present case, straight-line orbits of the dual system get mapped to circular orbits of the original problem.

The BAV duality has a long history that starts from a 1742 book by MacLaurin \cite[Book~II, Chapter~V, Section~875]{maclaurin_Fluxons1742}. His
result is described is the review \cite[Section~3]{albouy2022_253}. Historically, a map between a harmonic oscillator and a Coulomb problem has been the principal object of interest. The timeline of the development is presented in \cite[Section~10]{albouy2022_253}.

Note that the conventional BAV duality requires a conformal transformation from a Euclidean space onto itself, with the potential energy of the problem of interest playing the role of a metric tensor. However, the explanation for the
appearance of zero-energy off-center circular orbits in our case requires a generalization of the BAV scheme to a conformal map from a curved to an~Euclidean space, first introduced by Goursat in 1887 \cite{goursat1887_446}. The map turns out to be a stereographic projection \cite{rosenfeld_stereographic_projection1977}
from a sphere to its equatorial plane. Similarly to the inverse quartic case, zero-energy orbits of our potential map to a free motion orbits of the dual problem: those are represented by the grand circles on a sphere. Finally, according to a known property of the stereographic projection, these circles map back to the circular orbits of the potential we found.

The circularity of the orbits aside, the zero-energy orbits of \eqref{potential_720} are also closed curves. We identify a maximal superintegrability set of conserved quantities responsible for the phenomenon.

In the final section, we summarize our results, put them in a context, and suggest future projects. In particular, we propose to look at the case where
the potential center lies outside of a circular orbit; such orbits are conjectured to be generated by a free motion on a hyperbolic surface.

\section[Main result: the potential that generates off-center circular orbits]{Main result: the potential that generates off-center \\ circular orbits}

\begin{Theorem}[a potential description of the force \eqref{force_720}]\label{thm:main}
The expression \eqref{force_720} for the radial component of a central force that supports
a circular orbit \eqref{r_through_theta} admits a potential description with
\begin{align}
V(r) = -\frac{\alpha}{(r^2+\mathcal{R}^2)^2},\label{potential_720}
\end{align}
as the
potential, where $\mathcal{R} = \sqrt{R^2-l^2}$ is a half of the shortest chord of the orbit circle that crosses the
force center $($see Figure~{\rm \ref{f:main})}.
\end{Theorem}

\begin{proof}
According to the intersecting chords theorem, $r\times r' = \mathcal{R} \times \mathcal{R}$ (Figure~\ref{f:main}). On the
other hand, the radial component of the force produced by the potential \eqref{potential_720} is
\begin{align*}
\frac{\partial}{\partial r} V(r) = -\frac{4\alpha r}{\big(r^2+\mathcal{R}^2\big)^3}.
\end{align*}
Thus, Newton's expression for the force on the orbit, equation~\eqref{force_720}, can be transformed as
\begin{align*}
F= -\frac{4\alpha}{r^{2} \chord^{3}}
= -\frac{4\alpha}{r^{2} (r+r')^{3}}
= -\frac{4\alpha}{r^{2} \big(r+\frac{\mathcal{R}^2}{r}\big)^{3}}
= -\frac{4\alpha r}{\big(r^2+\mathcal{R}^2\big)^{3}}
= \frac{\partial}{\partial r} V(r).\tag*{\qed}
\end{align*}
\renewcommand{\qed}{}
\end{proof}

\begin{Remark}[orbits \eqref{r_through_theta} are zero-energy orbits]
The circular orbits \eqref{r_through_theta} are only a small subset of the orbits generated by the potential \eqref{potential_720}. In fact, they
are the zero-energy orbits for this potential. Indeed, consider the aphelion (the rightmost point of the orbit of Figure~\ref{f:main}) and perihelion
(the leftmost point there). Energy at these two points will read
\begin{align*}
E = \frac{L_{z}^2}{2m(R+l)^2} -\frac{\alpha}{\big((R+l)^2+\mathcal{R}^2\big)^2},\\
E = \frac{L_{z}^2}{2m(R-l)^2} -\frac{\alpha}{\big((R-l)^2+\mathcal{R}^2\big)^2},
\end{align*}
respectively,
where $L_{z} = m r^2 \dot{\phi} = \text{const}$ is the angular momentum. Solving for $E$ and $L_{z}$ we get
\begin{align*}
E=0,
\end{align*}
and $L_{z} = \pm \frac{1}{\sqrt{2}} \sqrt{\frac{m\alpha}{R^2}}$.

Conversely, every zero-energy orbit of the potential \eqref{potential_720} has a form \eqref{r_through_theta}, up to a rotation about the
potential center. This property follows from Remark~\ref{remark:geometry_of_orbits}.
\end{Remark}

\section[Why zero-energy orbits in a 1/(r\^{}2+R\^{}2)\^{}2 potential well are circular]{Why zero-energy orbits in a $\boldsymbol{1/\big(r^2+\mathcal{R}^2\big)^2}$ potential \\ well are circular}
The potential \eqref{potential_720} has been specifically constructed to support circular orbits at zero energy. Nonetheless,
one would want to know the deeper mathematical reason for this property. In this section, we will present an
example of such a reason.

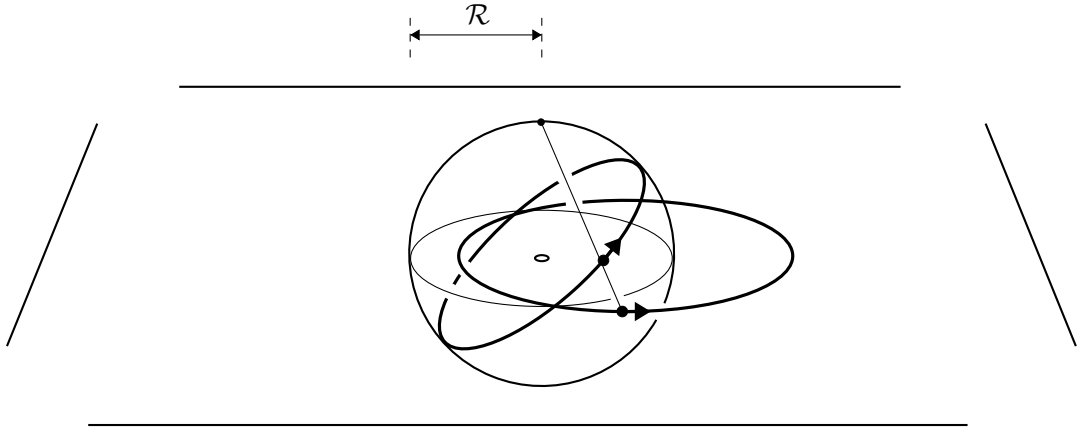
\begin{figure}[t]\centering
 \begin{tikzpicture}[scale=1,smooth]
 %\node at (0,0) {\includegraphics[scale=.5]{stereographic-eps-converted-to.pdf}};
 %\draw[very thin,color=gray,step=.5cm] (-7.5,-4.0) grid (7.5,4.0);
 %\draw[fill] (0,0) circle [radius=.1];
 \draw[thick] (-4.92,1.83) -- (4.67,1.83);
 \draw[thick] (-6.13,-2.67) -- (5.56,-2.67);
 \draw[thick] (-6.01,1.34) -- (-7.21,-1.62);
 \draw[thick] (5.8,1.34) -- (7,-1.62);
 \draw[Triangle-Triangle] (-0.1,2.52) -- (-1.85,2.52);
 \draw[thick] (1.53,-1.05) arc[start angle=-22, end angle=330, x radius=1.76cm, y radius =1.76cm];
 \draw[very thick,rotate=42] (1.4,-0.23) arc [start angle=0, end angle=360, x radius=1.74cm, y radius =0.64cm];
 \draw[fill,white] (-1.12,-0.59) circle [radius=.1];
 \draw[fill,white] (-1.32,-0.89) circle [radius=.1];
 \draw[fill,white] (0.2,0.63) circle [radius=.1];
 \draw[very thick] (0.22,0.27) arc [start angle=-249, end angle=105, x radius=2.22cm, y radius =0.74cm];
 \draw (0.99,-0.95) arc [start angle=-51, end angle=302, x radius=1.74cm, y radius =0.64cm];
 \draw[fill] (-0.11,1.36) circle [radius=.043];
 \draw[fill] (0.97,-1.16) circle [radius=.07];
 \draw[fill] (0.72,-0.48) circle [radius=.07];
 \draw[-Triangle,very thick] (0.97,-1.16) -- (1.35,-1.16);
 \draw[-Triangle,very thick] (0.72,-0.48) -- (0.96,-0.175);
 \draw (-0.11,1.36) -- (0.97,-1.16);
 \draw[thick] (-0.1,-0.45) ellipse (0.09 and 0.04);
 \node at (-0.95,2.8) {$\mathcal{R}$};
 \draw[thin,dashed] (-0.1,2.22) -- (-0.1,2.82);
 \draw[thin,dashed] (-1.85,2.22) -- (-1.85,2.82);
 \end{tikzpicture}
 \caption{A stereographic projection from a sphere of radius $\mathcal{R}$ to its equatorial plane.
 The spherical geodesics---the great circles---become, in the plane, circles that cross the sphere's equator at two diametrically opposite
 points. In the main text, we show that the free-motion orbits on the sphere become orbits in the potential \eqref{potential_720}.}\label{f:stereographic}
 \end{figure}

Let us embed the $xy$-plane we are working with in a three-dimensional space with coordinates $x$, $y$, and $z$,
and introduce a radius $\mathcal{R}$ sphere there with its center at the origin. Consider a~stereographic projection
from the sphere to the $xy$-plane of interest (Figure~\ref{f:stereographic}):
\begin{align}
\left(
\begin{matrix}
x
\\
y
\\
0
\end{matrix}
\right) = \frac{\bm{s}'-(\bm{e}_z \cdot \bm{s}')\bm{e}_z}{1-(\bm{e}_z \cdot \bm{s}')} \mathcal{R}.
\label{stereographic}
\end{align}
Its inverse reads
\begin{align}
s_{x}' = \frac{2 \mathcal{R}^2}{r^2 + \mathcal{R}^2} \frac{x}{\mathcal{R}}, \qquad
s_{y}' = \frac{2 \mathcal{R}^2}{r^2 + \mathcal{R}^2} \frac{y}{\mathcal{R}}, \qquad
s_{z}' = \frac{r^2 - \mathcal{R}^2} {r^2 + \mathcal{R}^2 }.\label{stereographic_inverse}
\end{align}
Here, $\bm{s}'$ is the unit vector directed to a point on the sphere and $\bm{e}_z$ is the unit vector in the $z$-direction.

\begin{Theorem}[correspondence between the motion in a $1/\big(r^2+\mathcal{R}^2\big)^2$ potential well and a~free motion on a~sphere]\label{thm:sphere}
The stereographic projection \eqref{stereographic} maps free motion orbits on the sphere~\eqref{stereographic_inverse} to the orbits in the
potential \eqref{potential_721}.
\end{Theorem}

\begin{proof}
An energy $E$ orbit $\Omega$ in the potential \eqref{potential_720} extremises
the Maupertuis--Jacobi functional
\begin{align*}
S[\mathcal{P}] = \int_{\mathcal{P}} \sqrt{2m (E-V(\bm{r}))} \, {\rm d}l,
\end{align*}
where $\mathcal{P}$ is any path in the coordinate space that shares the beginning and the end points with the orbit
\cite[p.~247]{arnold__methods_of_CM1989}. At zero energy, the
Maupertuis--Jacobi action becomes
\begin{align}
S[\mathcal{P}] = \sqrt{2m E'} \int_{\mathcal{P}}{\rm d}l',\label{sphere_distance}
\end{align}
with
\begin{align}
{\rm d}l' = \frac{2 \mathcal{R}^2}{r^2+\mathcal{R}^2} \, {\rm d}l
\qquad
\text{and}
\qquad
E' = \frac{\alpha}{4 \mathcal{R}^4}.\label{E_prime}
\end{align}
The length element ${\rm d}l'$ is the distance traveled on the sphere \eqref{stereographic_inverse} by the stereographic prototype of the point that traveled a distance ${\rm d}l$ in the $xy$-plane. The whole functional \eqref{sphere_distance} is then proportional
to the length of the prototype of the path $\mathcal{P}$ on the sphere. Thus, a path in $xy$ that extremises $\mathcal{P}$ (i.e., an actual orbit) will be a stereographic image of a free motion path on the prototype sphere.
\end{proof}

\begin{Remark}
Theorem \ref{thm:sphere} is a particular case of a much more general result that covers any conformal map between two Riemannian
manifolds, at finite as well as at zero energy \cite{goursat1887_446}.
\end{Remark}

\begin{Remark}[why orbits are circles]
A stereographic projection from a sphere to its equatorial plane
maps circles (that do not cross the north pole) on the sphere to circles in the equatorial plane \cite{rosenfeld_stereographic_projection1977}. This explains why the
zero energy orbits of \eqref{potential_720} are circles.
\end{Remark}

\begin{figure}[t]\centering
 \begin{tikzpicture}[scale=1,smooth]
 %\node at (0,0) {\includegraphics[scale=.3]{orbits.pdf}};
 % \draw[very thin,color=gray,step=.5cm] (-7.0,-5.0) grid (7.0,5.0);
 % \draw[fill,black] (0,0) circle [radius=.1];
 \draw [thick] (0.45,0.25) circle [radius=3.25];
 \draw [thick] (-0.55,-0.2) circle [radius=2.45];
 \draw [thick] (-1.57,-0.68) circle [radius=1.95];
 \draw [thick,dashed] (-1.95,-0.85) circle [radius=1.87];
 \draw [thin,dashed] (-2.75,0.9) -- (-1.13,-2.55);
 \draw [ultra thick,o-Triangle] (-2.07,-0.9) -- (-1.05,-0.45);
 %\draw [ultra thick,-Stealth] (-2.07,-0.9) -- (-1.05,-0.45);
 \node at (-1.6,-0.4) {$\vec{n}$};
 \node at (-3.75,-2) {$\mathcal{R}$};
 \end{tikzpicture}
 \caption{Three representative zero-energy orbits in the potential \eqref{potential_720}, for a given displacement orientation $\bm{n}$ (solid lines).
The origin-centered circle of radius $\mathcal{R}$ (dashed-line) is the intersection of the stereographic sphere with the coordinate plane. The thin dotted line segment is circle's diameter.}\label{f:orbits}
 \end{figure}
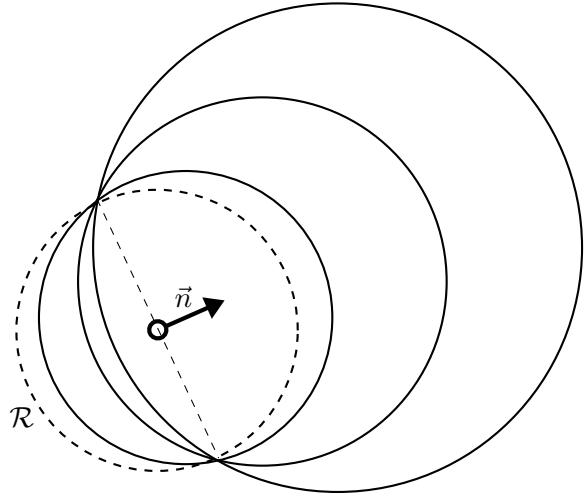

\begin{Remark}[geometry of the orbits, for a given potential]\label{remark:geometry_of_orbits}
For a given potential \eqref{potential_720}, the family of zero-energy orbits consists of circles that cross an origin-centered circle
of a radius $\mathcal{R}$ at two points diametrically opposite to each other (see Figure~\ref{f:orbits}). Note that
this radius $\mathcal{R}$ circle is the intersection between the $xy$-plane and the stereographic sphere.
\end{Remark}

\begin{Remark}[relationship between \emph{trajectories}]
Consider a \emph{trajectory} of a free particle of an energy $E'$ given by \eqref{E_prime}, on the sphere \eqref{stereographic_inverse}.
One can show that $\bm{r}(\bm{s}'(t'(t)))$ will be a valid zero-energy trajectory in the potential \eqref{potential_720}, where the
inverse stereographic map $\bm{r}(\bm{s}')$ is given by \eqref{stereographic_inverse} and $t'(t)$ obeys
\begin{align}
\frac{{\rm d}t'}{{\rm d}t} = \frac{4 \mathcal{R}^4}{\big(r(t)^2+\mathcal{R}^2\big)^2}.\label{720_time_dilation}
\end{align}
\end{Remark}

\begin{Remark}[the known limit of a $1/r^4$ potential well]
The well-studied $r \gg \mathcal{R}$ limit leads to the inverse quartic potential \eqref{potential_721}. There, the inverse stereographic
projection \eqref{stereographic_inverse} degenerates onto a circle inversion, with
\begin{align*}
\mathcal{R}_{0} = \sqrt{2} \mathcal{R}
\end{align*}
being the radius of the inversion circle, itself centered at the origin.
\end{Remark}

\section[Why zero-energy orbits in a 1/(r\^{}2+R\^{}2)\^{}2 are closed: an additional integral of motion]{Why zero-energy orbits in a $\boldsymbol{1/\big(r^2+\mathcal{R}^2\big)^2}$ are closed: \\ an additional integral of motion}

Circular or not, the zero-energy orbits of our model are also closed. Such a phenomenon is usually associated with a presence of
$2N - 1$ functionally independent integrals of motion where $N$ is the number of degrees of freedom (the so-called maximal super-integrability set). In our two-dimensional system, we
need to find one additional constant of motion besides the Hamiltonian
\begin{align*}
H(\bm{r},\bm{p}) = \frac{p^2}{2m} - \frac{\alpha}{\big(r^2+\mathcal{R}^2\big)^2}
\end{align*}
and the angular momentum
\begin{align*}
L_{z}(\bm{r},\bm{p}) = x p_{y} - y p_{x} .
\end{align*}
Here, $\bm{p} = m \frac{{\rm d}}{{\rm d}t} \bm{r}$ is the momentum vector.
\begin{Theorem}[an additional integral of motion]\label{thm:additional_integral}
At zero energy, a unit vector
\begin{align}
\bm{n} \equiv -\sign(L_{z}) \frac{\bm{I}_{xy}}{|\bm{I}_{xy}|},\label{integrals_unit_vector}
\end{align}
with
\begin{align}
\bm{I}_{xy} \equiv
\left(
\begin{matrix}
I_{x}(\bm{r},\bm{p})
\\
I_{y}(\bm{r},\bm{p})
\end{matrix}
\right)
= -\frac{1}{2\mathcal{R}}
\big(
\bm{r} L_{z}(\bm{r},\bm{p}) +
(\bm{e}_{z} \times \bm{r})
Q(\bm{r},\bm{p}) +
(\bm{e}_{z} \times \bm{p})
\mathcal{R}^2
\big),
\label{integrals}
\end{align}
is conserved, and it is functionally independent from both the Hamiltonian and the angular momentum. Here,
\[Q(\bm{r},\bm{p}) = \bm{r} \cdot \bm{p}.\]
\end{Theorem}

\begin{proof}
A straightforward calculation shows that
\begin{align*}
\left\{\bm{I}_{xy},H\right\} = -\frac{2}{\mathcal{R}}
\left(
\begin{matrix}
-y
\\
+x
\end{matrix}
\right)
H;
\end{align*}
here and below, $\{A,B\}
\equiv \frac{\partial A}{\partial \bm{r}} \frac{\partial B}{\partial \bm{p}} - \frac{\partial A}{\partial \bm{p}} \frac{\partial B}{\partial \bm{r}}$
are the Poisson brackets. Indeed, this commutator vanishes at zero energy.

For an off-center zero-energy orbit, $|\bm{I}_{xy}|$ is strictly greater than zero: indeed, one can show that the
difference between the aphelion ($r_{\max}$) and perihelion ($r_{\min}$) distances is
\begin{align*}
r_{\max} - r_{\min} = 2\sqrt{\frac{|\bm{I}_{xy}|}{|L_{z}|}} \mathcal{R}.
\end{align*}
Consider such an orbit. If one rotates it about the origin, both the energy $H$ and the angular momentum $L_{z}$
will remain unchanged, but the orientation of the vector $\bm{I}_{xy}$ changes together with the orbit. This proves
the functional independence of $\bm{n}$ of $H$ and $L_{z}$, only for the off-center orbits.

In the case of a circular orbit centered at the origin, the functional independence of $\bm{n}$ is not needed for
a proof of the orbit closeness.
\end{proof}

\begin{Remark}[magnitude of $\bm{I}_{xy}$]
The magnitude of the vector $\bm{I}_{xy}$,
\begin{align*}
|\bm{I}_{xy}| = \sqrt{\frac{m \alpha}{2 \mathcal{R}^2} - L_{z}^2},
\end{align*}
is a function of $L_{z}$.
\end{Remark}

\begin{Remark}[the physical meaning of the integrals of motion \eqref{integrals}]
Note that each of the three components of the angular momentum on the sphere,
\begin{align*}
I'_{i} = m \mathcal{R}^2 \left(\bm{s}' \times \frac{{\rm d}}{{\rm d}t'}\bm{s}' \right)_{i},
\end{align*}
is conserved. Here, $i=1,2,3$ correspond to $x$, $y$, $z$, with a cyclic convention
$2+2\stackrel{\cdot}{=}1$, $3+1\stackrel{\cdot}{=}1$, $3+2\stackrel{\cdot}{=}2$. One can now express these conserved quantities
through the variables of the original problem, using the connections \eqref{stereographic_inverse} and \eqref{720_time_dilation}.
It can be shown that the 2D $I_{x,y}(t) = I'_{x,y}(t'(t))$, while
the 2D angular momentum $L_{z}=I'_{z}$.
\end{Remark}

\begin{Remark}[the analogy with the Runge--Lenz vector]
The vector $\bm{I}_{xy}$, together with the angular momentum $L_{z}$ controls the displacement
of the orbit center from the force center:
\begin{align*}
l \bm{n} = -\frac{\mathcal{R}}{L_{z}} \bm{I}_{xy}.
\end{align*}
The vector $\bm{n}$ will determines the orientation of the position of the orbit center, $l \bm{n}$, with
respect to the origin; as such, it
is analogous to the Laplace–Runge–Lenz vector of the
Coulomb problem.
\end{Remark}

\section{Summary, interpretation of the results, and outlook}
This article is inspired by Proposition 7, Problem 2 of Book 1 of Newton's \emph{Principia}. There, Newton gives a formula for how a force should depend on the point \emph{on the orbit} in order for the orbit to be an \emph{off-center} circle. We identify
a potentialthat can induce such a force~\eqref{potential_720}. All \emph{zero-energy} orbits in this potential are, indeed, off-center circles.

Next, we show that the motion in the potential \eqref{potential_720} is a stereographic image of free motion on a sphere.
This connection can be regarded as a manifestation of the Bohlin--Arnold--Vassiliev (BAV) duality \cite{arnold_HBNH_1990,arnold1990_1148,bohlin1911_113,horvathy2014_14042265,maclaurin_Fluxons1742} introduced by MacLaurin, in 1742, and its non-Euclidean generalization by Goursat \cite{goursat1887_446}, from 1887. At large distances from the potential center, where our potential becomes an inverse quartic well, our map degenerates to a circle inversion; the latter relates the motion in the inverse quartic potential to a free motion in a flat space, a well-known manifestation of the BAV duality \cite{arnold_HBNH_1990} (a self-duality, in this particular case).

Since the zero-energy orbits
in the potential \eqref{potential_720} are closed curves, we were also compelled to identify an additional integral of motion,
complementary to the Hamiltonian and the angular momentum. This integral is
an analogue of the Runge--Lenz vector of the Coulomb problem.

For future projects, it appears promising to search for other problems that map to a free motion on curved surfaces with solvable geodesics.

In particular, Newton's result can be generalized to the case where the potential center lies
\emph{outside} of the orbit. This motion leads to a potential
\begin{align*}
V_{\text{hyperbolic}}(r) = -\frac{\alpha}{\big(r^2-\tilde{\mathcal{R}}^2\big)^2}
,\qquad \text{with}\quad \tilde{\mathcal{R}} = \sqrt{l^2-R^2},
\end{align*}
where $R$ and $l$ are the radius of the orbit and the distance between its center and the force center respectively.
We expect that this potential would induce
a metric of the Poincar\'{e} disk model of hyperbolic geometry \cite[p.~45]{penrose_road_to_reality2004}. There, the geodesics
are circle segments intersecting the border of the Poincar\'{e} disk at a right angle, inside the disk. We would like to point out
that in this case, it can be predicted that the analogue of the integral of motion \eqref{integrals_unit_vector} will only
be conserved in between the subsequent collisions with the singularity at $r=\tilde{\mathcal{R}}$; angular momentum conservation
is likely to require sudden jumps in $\bm{n}$, at the moment of such a collision.

\subsection*{Acknowledgements}
We thank Francisco J.\ Jauffred and Bala Sundaram for inspiring this project and Vanja Dunjko and Steven Jackson for the helpful
discussions. We are indebted to the anonymous referee who inspired us to drastically reorganize the original draft and provided
the absolutely crucial references \cite{maclaurin_Fluxons1742} and \cite{goursat1887_446}. This work was supported by NSF grant PHY-1912542.

\pdfbookmark[1]{References}{ref}
\LastPageEnding

\end{document}